\documentclass[11pt]{article}
\usepackage{graphicx}

\newsavebox{\hflrar}
\sbox{\hflrar}{\makebox[0pt][l]
{${\scriptstyle \leftharpoonup}$}{${\scriptstyle \rightharpoonup}$}}

\def \to {\rightarrow}

\begin{document}
\begin{center}
{\Large\bf Resummation of Large Logarithms in the Electromagnetic Form
factor of $\pi$ }
\vskip 10mm
F. Feng$^{1}$ and J.P. Ma$^{2,3}$     \\
{\small {\it $^1$ Theoretical Physics Center for Science Facilities, Institute of High Energy Physics,
Academia Sinica,
Beijing 100049, China }} \\
{\small {\it $^2$ Institute of Theoretical Physics, Academia Sinica,
Beijing 100190, China }} \\
{\small {\it $^3$ Institute of Particle Physics and Cosmology, Department of Physics,
Shanghai Jiao Tong University, Shanghai 200240,China }} \\
\end{center}

\vskip 1cm
\begin{abstract}
In the collinear factorization of the form factor for the transition
$\gamma^* \pi^0 \to \pi$  the hard part contains double log terms.
These terms will spoil the perturbative expansion of the hard part.
A simple exponentiation for resummation leads to divergent
results. We study the resummation of these double log's. We make an analysis
to show the origin of the double log's.
With the understanding of the origin one can introduce soft factors and
Nonstandard Light-Cone Wave Functions(NLCWF) to derive a factorized form
for the form factor, where the hard part does not contain the double log's.
There is a perturbative relation between NLCWF and the standard Light-Cone
Wave Function(LCWF).
Beside the renormalization scale $\mu$ the introduced NLCWF's
and soft factors
have extra scales to characterize the double log's.
Using the evolutions
of the extra scales and the relation we can do the resummation of the double log's
perturbatively in sense that LCWF's are the only nonpertubative objects
in the resumed formula. Our results with some models of LCWF
show that there is a significant difference between numerical predictions with or without the resummation.
\vskip 5mm \noindent
\end{abstract}
\vskip 1cm
\par\vfil
\eject

\noindent
{\bf\large 1. Introduction}
\par\vskip15pt
Although the interaction of QCD is asymptotically weak , the perturbative theory of QCD
can not directly be used
to study hadronic processes involving large momentum transfers because of quark confinement.
One needs
to separate or factorize long-distance- and short-distance effects. Only the latter,
which are characterized by large momentum transfers denoted generically as $Q$
can be studied with perturbative QCD.
For an exclusive process
it has been proposed long time ago that
such a process can be studied by an expansion of
the amplitude in $1/Q$, corresponding to an expansion of QCD
operators in twist\cite{BL,CZrep}. The leading term can be
factorized as a convolution of a perturbative coefficient function and light-cone wave
functions of hadrons. The light-cone wave functions are defined with
QCD operators. The perturbative coefficient function, which can be safely calculated
with perturbative QCD, describes hard scattering of
partons at short distances.
This is the so-called collinear
factorization. In this factorization the transverse
momenta of partons in parent hadrons are also expanded in the hard
scattering part and they are neglected at leading twist.
\par
Perturbative coefficient functions in collinear factorization contain
large logarithms at higher orders of
$\alpha_s$. These large logarithms are dangerous and can spoil the expansion
in the sense that the expansion does not converge. A resummation
of large logarithms
is often needed to have a reliable prediction. In this paper we study
the resummation in the process $\gamma^* \pi^+ \to \pi^+$. The process is described
by the electromagnetic form factor $F(Q)$ with $Q$ as the virtuality of the virtual photon.
For large $Q$ the form factor can be written as a convolution
\begin{equation}
 F(Q) \sim \phi (x) \otimes \phi(y)\otimes H(x,y)\left [ 1 + {\mathcal O}(\Lambda^2/Q^2) \right ]
\end{equation}
with $\Lambda$ as a scale characterizing nonperturbative effects.
In the above  $\phi(x)$ is the light-cone wave function(LCWF) of $\pi$ with the momentum fraction
$x$ carried by a parton,
$H$ is the hard part which can be calculated safely as an expansion in $\alpha_s$.
The one-loop correction to the hard part has been studied in \cite{FGOC,DR,KR,BT,BNP}.
It has been found that the hard part at one-loop contains double log terms like
$\alpha_s^2 \ln^2 x$ and $\alpha_s^2 \ln^2 y$. At $n$-loop level
the hard part will contain $ \alpha_s^{1 +n} \ln^{2n} x$ or $ \alpha_s^{1 +n} \ln^{2n} y$.
Those double log terms will become divergent when $x$ or $y$ is approaching to zero
and can spoil the perturbative
expansion of $H$. Although the double log terms are integrable with $x$ or $y$ in the convolution
and give finite contributions, but they are significant corrections.
The purpose of our study is to resum those double log terms.
\par
We  study the resummation of the above double log terms in the collinear factorization
by re-factorizing the double log terms in the hard part. For doing this we have to understand
the origin of the double log.
Given the factorized form in the above, $H$ will receive
in general contributions from the form factor and the LCWF.
If we use a
finite quark mass to regularize the collinear singularities, we can
show that a part of  double log terms come from LCWF.
This part can be re-factorized by using a nonstandard light-cone wave function(NLCWF)
which has a perturbative relation to LCWF.
The remaining double log terms come from the form factor. They can be re-factorized
into soft factors.
In introducing NLCWF and soft factors non-light like gauge
links play an important role. The importance of using non-light like gauge
links has been shown in Transverse-Momentum-Dependent(TMD) factorization
for inclusive processes\cite{CS,CSS,JMY,JMYG}.
With these gauge links extra scales
are introduced to control these double log terms.
Using evolution equations of these scales we are able to resum the above
double log terms. The outlined approach has been used to resum double log terms
in $\pi \gamma^* \to \gamma$ in \cite{FMW}, where the resummation has a significant
effect. Our approach is similar to the threshold resummation in
inclusive processes studied in \cite{Ster}.
\par
The resummation of these double log terms has been studied in \cite{LS}
with the $k_T$-factorization. The $k_T$-factorization has been widely used in studies
of $B$-meson decays. However, such a factorization is not gauge-invariant because
hard parts are extracted from scattering amplitudes of off-shell partons. The amplitudes
of off-shell partons
are in general gauge-dependent. Recently it has been shown that the hard parts
in the factorization receive at loop-level divergent contributions which are gauge-dependent\cite{VKT}.
One may find further
discussions of the issue in \cite{FVKT}. The $k_T$-factorization takes transverse momenta of partons
into account. These momenta are neglected at leading twist in the collinear factorization.
It should be noted that the effects of the transverse momenta can be taken into account
in a gauge-invariant way by using the so-called Transverse Momentum Dependent(TMD) factorization\cite{TMD1,TMDB}.
It is possible to use the TMD factorization to perform the resummation. But it is at moment
not possible to give numerical predictions in detail because the involved nonperturbative objects
are poorly known.
\par
Our paper is organized as the following: In Sec. 2 we introduce our notations
and give a brief discussion about the consequence of the double log terms from higher orders.
In Sec. 3 we explain the origin of double log terms in the hard part.
In Sec. 4 we introduce NLCWF and soft-factors to re-factorize double log terms.
In Sec. 5 we show that these double log terms can be resummed and give our numerical
results. Sec. 6 is our conclusion. An appendix is given to discuss the problem
of gauge invariance in $k_T$-factorization for the case studied here.

\par\vskip20pt
\noindent
{\bf 2. Notations}
\par\vskip15pt\noindent
\par
The electromagnetic form factor of $\pi^+$ is defined as:
\begin{equation}
\langle \pi^+ (K) \vert J^\mu \vert \pi^+(P) \rangle = F_\pi (Q) (P+K)^\mu.
\end{equation}
We will use the  light-cone coordinate system, in which a
vector $a^\mu$ is expressed as $a^\mu = (a^+, a^-, \vec a_\perp) =
((a^0+a^3)/\sqrt{2}, (a^0-a^3)/\sqrt{2}, a^1, a^2)$ and $a_\perp^2
=(a^1)^2+(a^2)^2$. Two vectors $l^\mu =(1,0,0,0)$ and $n^\mu=(0,1,0,0)$ are introduced.
We take a light-cone coordinate system in which the momenta are given as:
\begin{eqnarray}
P^\mu  \approx (P^+,0,0,0),
\ \ \ \ K^\mu \approx (0,K^-,0,0),\ \ \
 q^\mu &=& K-P, \ \ \ \ Q^2 =-q^2 \approx 2P^+ K^-.
\end{eqnarray}
When $Q^2$ is very large, the form factor takes a factorized form. To derive
the factorized form, i.e., to determine the hard part,
one usually replaces hadronic state with partonic states to calculate
the form factor and LCWF's.
We replace the initial state $\vert \pi^+ (P) \rangle$ with the partonic state
$\vert \bar d(p_1) u(p_2) \rangle$,
and the final state $\langle \pi^+ (K) \vert $
with the partonic state $\langle  u(k_2) \bar d(k_1) \vert $.
The quark pairs are in color-singlet.
The momenta are:
\begin{eqnarray}
p_1^\mu =\left  (\bar x_0 P^+,0,0,0 \right ), \ \ \
p_2^\mu = \left  ( x_0 P^+, 0, 0,0 \right ),
\ \ \
k_1^\mu = \left  ( 0, \bar y_0 K^-, 0,0  \right ), \ \ \
\ k_2^\mu = \left  (0, y_0 K^-, 0,0  \right )
\end{eqnarray}
with $\bar x_0 =1-x_0$ and $\bar y_0 =1-y_0$.
We take massless quarks and all singularities are regularized with dimensional regularization.
\par

\begin{figure}[hbt]
\begin{center}
\includegraphics[width=9cm]{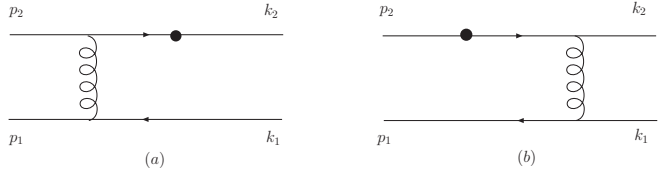}
\end{center}
\caption{The leading order contributions from the $u$-quark. The
black dot denotes the insertion of the electromagnetic current, i.e.,
inserting a $\gamma^\mu$.} \label{Feynman-dg1}
\end{figure}
\par
The contributions to the form factor of the replaced partonic states with the virtual
photon attached to the $u$-quark are from two
diagrams given in Fig.1.
To reduce the number of diagrams which need to be calculated, we take $\mu= -$.
Fig.1b will not contribute for $\mu= -$.
It is straightforward to obtain the $u$-quark contribution:
\begin{eqnarray}
 F_\pi (Q)\biggr \vert_{1a} =   2 e_u \frac{g_s^2}{\bar x_0 \bar y_0 Q^4}  \bar v(p_1) \gamma^\rho T^a v(k_1)
                 \bar u(k_2)  \gamma_\rho T^a u(p_2),
\end{eqnarray}
where $e_u$ is the electric charge of the $u$-quark. Similarly, one can obtain the contribution from
the $\bar d$-quark through the symmetry of charge conjugation.
\par
The definition of LCWF is:
\begin{eqnarray}
\phi(x, \mu) &=& \ \int \frac{ d z^- }{2\pi}
    e^{i x P^+ z^- }
\langle 0 \vert \bar d (0) L_n^\dagger (\infty, 0)
  \gamma^+ \gamma_5 L_n (\infty,z) u (z^- n) \vert \pi^+(P) \rangle ,
\end{eqnarray}
where $q(x)(q=\bar d, u) $ is the light-quark field. $L_n$ is the gauge
link in the direction $n$:
\begin{equation}
L_n (\infty, z) = P \exp \left ( -i g_s \int_{0} ^{\infty} d\lambda
     n \cdot G (\lambda n+ z ) \right ) .
\end{equation}
If we replace $\pi$ with the parton state we have at the tree level for the wave function:
\begin{equation}
\phi^{(0)} (x, \mu) =  \delta (x-x_0) \phi_0 + \cdots, \ \ \ \ \  \phi_0 =\bar
 v(p_1) \gamma^+ \gamma_5 u(p_2) /P^+,
\end{equation}
where $\cdots$ stand for the states of the quark pair with quantum numbers other than that
of $\pi$.
For the outgoing $\pi^+$ LCWF is defined with quark fields which are separated in the direction $l$
and  we the corresponding gauge link is along the direction of $l$.
With the definitions one easily obtains the factorized form:
\begin{eqnarray}
F_\pi (Q) & \approx &
   \frac{8 \pi \alpha_s}{9 Q^2}
\int_0^1 dx dy  \phi (x,\mu)
\phi  (y,\mu)
 \cdot \frac{1}{ \bar x  \bar y} {\mathcal H}(x,y,Q,\mu)
\nonumber\\
{\mathcal H}(x,y,Q,\mu) &=& 1 + {\mathcal O} (\alpha_s).
\end{eqnarray}
In the above the hard part ${\mathcal H}$ can be calculated as an expansion in $\alpha_s$.
The correction to the factorized form factor is power-suppressed.
\par
At one-loop level the hard part receives corrections which have double log terms\cite{FGOC,DR,KR,BT,BNP}:
\begin{equation}
{\mathcal H}(x,y,Q,\mu) = 1 + \frac{\alpha_s}{3 \pi} \left [ \ln^2 \bar x + \ln^2 \bar y \right ] + \cdots,
\end{equation}
and at higher orders of $\alpha_s^n$ those terms like $ \alpha^n_s \ln^{2n} \bar x$ and  $\alpha_s^n \ln^{2n} \bar y$,
which can lead to an divergent series in $\alpha_s$. If we take asymptotic form of LCWF, i.e.,
$\phi(x) \sim x(1-x)$, one easily finds that the $1+n$-th order contribution to the form factor:
\begin{equation}
 F^{(1+n)} (Q) \sim \alpha_s^{1+n} (2n)!.
\end{equation}
The factorial increase of the $1+n$-th order contribution can spoil the perturbative expansion.
Therefore, these double log terms need to be resummed in order to have
a reliable expansion. It should be noted here that a simple exponentiation of these double log's leads
to a divergent form factor because the coefficient in the front of the double log's is positive.
\par
For the $k_T$-factorization in the case studied here the same diagrams in Fig.1 gives the leading-order
result, in which the partons are off-shell. Therefore, the result is not gauge-invariant.
We will discuss this in detail in Appendix.

\par
\vskip20pt
\par\noindent
{\bf 3. The Origin of the Double Log Terms}
\par\vskip15pt
\par
In this section we analyze the origin of the double log terms in the hard part.
The one-loop correction has been calculated in \cite{FGOC,DR,KR,BT,BNP}. In \cite{FGOC,BNP}
the individual contributions from each one-loop diagram are listed in detail. From their results
we find with $\mu=-$ in Eq. (2) that the diagrams given in Fig.2 give contributions containing double log's.
In order to have short notations, we denote in this section the momentum fraction $x_0$ and $y_0$ in Eq.(4)
as $x$ and $y$, respectively.
\par
In general the double log terms are
generated if the integration region of loop momenta has an overlap between collinear- and infrared region.
The infrared region always exists.
The collinear region can exist in two cases. One case is that a loop momentum $k$
is directly collinear to an external on-shell particle. In this case the integration of $k$
will generate a collinear singularity regularized with a mass of the external particle
or with dimensional regularization. Another case
is that in some kinematical region. e.g. when $\bar x$ or $\bar y$ approaches zero,
some propagators can become nearly on-shell and the collinear region is formed when $k$
is collinear to the momenta of these propagators. The integration over $k$
will generate the corresponding collinear singularities appearing in the form as $\ln \bar x$ or $\ln\bar y$.
If there is an overlap of these collinear regions with the infrared region, double log terms
will be generated. We will call those double log terms in the former case as
Type I double log's, and these in the later case as Type II double log's.
The existence of the two types of double log's will be explained in detail with the examples
in Fig.2.
\par

\begin{figure}[hbt]
\begin{center}
\includegraphics[width=12cm]{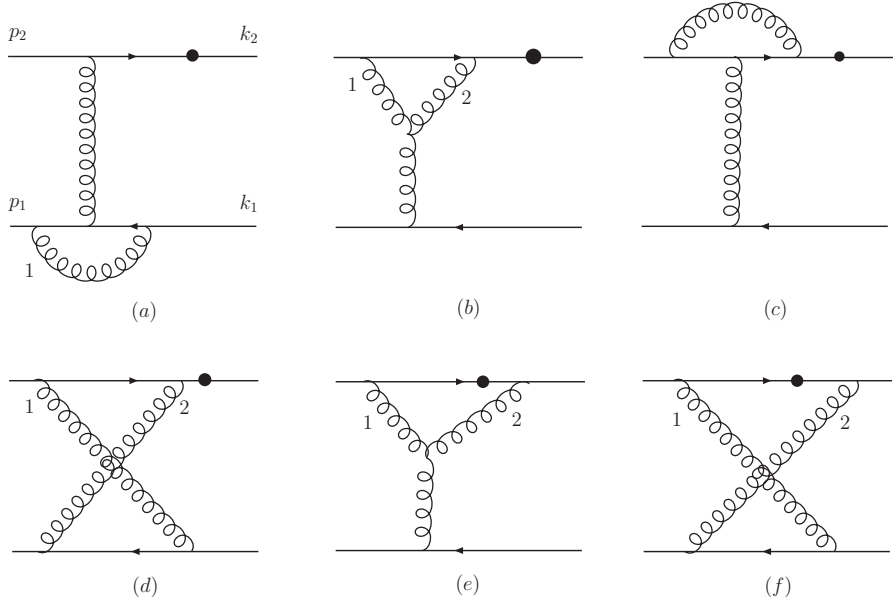}
\end{center}
\caption{One-loop correction with double log terms, the original labels of the diagrams
in \cite{BNP} are also given.}
\label{Feynman-dg2}
\end{figure}
\par
Let us first consider the contribution from Fig.2a.
This contribution contains double-log terms like $\ln^2\bar x$ and $\ln^2\bar y$.
It also contains a double pole term. These terms come from the overlap region of collinear- and soft gluons.
The double-pole and
double-log terms can easily be found by the approach of soft gluons where one neglects in the first step
all loop momenta $k$ in the nominator and picks the dominant terms in the denominator for
$k\to 0$. Then we have the soft part from Fig.2a:
\begin{eqnarray}
 F_\pi (Q) \biggr\vert_{2a}
  &\approx & F_\pi^{(0)} (Q)  \cdot \left ( \frac{ ig_s^2}{2N_c} \right ) \int \frac{ d^4 k}{(2\pi)^4}
\frac{4 p_1\cdot k_1 }{ ((p_1+k)^2+i\varepsilon) ((k_1 +k)^2+i\varepsilon)
 (k^2 +i\varepsilon)},
\nonumber\\
  &=&  F_\pi^{(0)} (Q) \left [ \frac{\alpha_s}{4\pi N_c} \left ( \left ( \frac{2}{\epsilon} \right )^2
     + \frac{2}{\epsilon}\left ( -\gamma + \ln \frac{4\pi \mu^2}{\bar x \bar y Q^2 }\right )
        + \frac{1}{2} \ln^2 \frac{4\pi \mu^2}{\bar x \bar y Q^2 } \right ) \right ].
\end{eqnarray}
\par
The reason for the double log's and the double pole in Fig.2a is the following: The exchanged gluon 1 can be collinear to
the initial $\bar q$ with the momentum $p_1$, it can also be soft. This overlapped region produces
an double pole and the double log of $p_1^+$, i.e., $\ln^2\bar x$. Similarly,
the gluon can also be collinear to the antiquark
in the final state with the momentum $k_1$. Again there is an overlapped region of the soft momentum.
This produces another double pole and the double log of $k_1^-$, i.e., $\ln^2\bar y$.
The sum of the two double pole contributions is given in the above. It is clearly that this type
of double log's is associated with partons in the initial- or final state. According to the discussion at the beginning
of this section, these double log's are Type I double log's.
Actually, the double log's and the double pole contributions from Fig.2a
are canceled
in the final result because the quark pair in the initial- and final state is neutral in color.
We note that the soft gluon emitted by the initial $\bar q$ can be absorbed
by the final $\bar q$ and the final $q$ quark, the former contribution
is given by Fig.2a, the later contribution is given by Fig.2f when the gluon 2 is soft.
The sum of the two contributions represents the interaction of the soft gluon
with the quark pair in the final state.
The interaction of the soft gluon  with the quark pair vanishes because
the pair is color-neutral.
Therefore, the double pole and double log associated
with $p_1$ in Fig.2a are canceled by those in Fig.2f.
With the same reason, one finds that the double pole and double log associated
with $k_1$ in Fig.2a are canceled by the soft contribution from Fig.2d with the gluon 1 as the soft
gluon. We will show this explicitly.
\par
We turn to the contribution from Fig.2b. In Fig.2b any one of the two gluons from
the gluon splitting can be soft. We denote the momentum of the gluon 1 as $k$.
The gluon 2 carries the momentum
$(k_1-p_1-k)$. If the gluon $1$ is soft, we have the approximation for $\bar x \to 0$:
\begin{eqnarray}
 F_\pi (Q^2) \vert_{2b} &=&  \frac{g_s^4 f^{abc}} {K^- (p_1-k_1)^2 (P-k_1)^2}
 \int \frac{d^4 k}{(2\pi)^4} \frac{ 8 k_1^- k_1 \cdot p_2}{(k^2 +i\varepsilon) ((k_1-p_1-k)^2 +i\varepsilon)}
\nonumber\\
   && \cdot \frac{1}{ -2 p_2^+ k^- +i\varepsilon} \left [ \bar u(k_2) \gamma^\alpha T^c  T^b u(p_2)
        \bar v(p_1) \gamma^\alpha T^a v(k_1) \right ] +\cdots
\nonumber\\
    &=&  F_\pi^{(0)} (Q^2) \left \{ \frac{ \alpha_s N_c}{4\pi} \left [ \left ( -\frac{2}{\epsilon}
         +\gamma  +\ln\frac{ \bar x \bar y Q^2}{4\pi\mu^2} \right ) \ln \bar x  -\frac{1}{2} \ln^2 \bar x \right ] \right\}
         +\cdots.
\end{eqnarray}
The term $1/( -2 p_2^+ k^- +i\varepsilon)$ is the quark propagator with the soft gluon approximation.
This contribution only generates the double log of $\bar x$. It also contains a collinear singularity
from the region in which $k$ is collinear to $p_2$.
For the case that the gluon 2 is soft, no double log term is generated.
There is also no double log term $\ln^2\bar y$ for $\bar y \to 0$.
For our purpose it is important to understand the above double log terms.
\par
In Eq.(13) there are two terms with the double log of $p_1^+$. One is associated with a collinear divergence,
one is given as $- (\ln^2 \bar x)/2$. These two terms have different origins. The initial quark emits a collinear
gluon and the gluon is absorbed by the virtual gluon attached to the antiquark line.
The virtual gluon becomes on-shell when $\bar x$ approaches to $0$.
With $\bar x =0$ one will find a double pole, one is collinear, another is I.R. one. In the case
with $\bar x \neq  0$, the I.R. singularity is regularized and it generates one $\ln\bar x$. Another
$\ln\bar x$ is generated from the collinear region. This is the origin for the first double log.
It is clear that this double log is associated with the collinear region of $p_2$ and this double
log is a Type I double log. The existence of Type I
double log also depends
on how the collinear singularity is regularized.
To see this more clearly, we can use a quark mass $m$
to regularize the collinear singularity. Then Eq.(13) becomes for $\bar x \to 0$:
\begin{eqnarray}
 F_\pi (Q^2) \vert_{2b} &=&  \frac{g_s^4 f^{abc}} { K^-(p_1-k_1)^2 (P-k_1)^2}
 \int \frac{d^4 k}{(2\pi)^4} \frac{ 8 k_1^- k_1 \cdot p_2}{(k^2 +i\varepsilon)((k_1-p_1-k)^2 +i\varepsilon)}
\nonumber\\
   && \cdot  \frac{1}{ (k-p_2)^2- m^2 +i\varepsilon}\left [ \bar u(k_2) \gamma^\alpha T^c  T^b u(p_2)
        \bar v(p_1) \gamma^\alpha T^a v(k_1) \right ] +\cdots
\nonumber\\
    &=&  F_\pi^{(0)} (Q^2) \left \{ \frac{ \alpha_s N_c}{4\pi} \left [
      -\ln \frac{m^2}{\bar y Q^2} \ln \bar x
      -\frac{1}{2} \ln^2 \bar x +\cdots \right] +\cdots \right \} .
\end{eqnarray}
From the above, one can see that the discussed double log disappear with the quark mass.
It should be noted that the hard part does not depend
how the collinear singularities are regularized. If one uses the mass
to regularize the collinear singularity,
one should also calculate the wave functions with the mass in order to subtract
collinear singularities in the scattering amplitude. Then
the Type I
of double log's appearing in the hard part come from the wave functions.
As we will show that Type I double log's can be factorized with NLCWF introduced
in \cite{FMW}.
\par
The second double log in Eq.(13), or that in Eq.(14) has a different origin
than that of Type I double log's. The virtual gluon attached with the antiquark line
carries the momentum $p_1-k_1$. For $\bar x\to 0$, the virtual gluon approaches on-shell,
its momentum is nearly collinear to $-k_1$. Here again
the gluon 1 can have a momentum region which is the overlap region with the loop momentum
is collinear to $-k_1$ and soft. For small non-vanishing $\bar x$ the collinear singularity
and the I.R. singularity are simultaneously regularized by $\bar x$, and it results in
the second double log $\ln^2\bar x$. This double log is a Type II double log.
As we will show later, this type of double log's can be factorized by certain soft factors.
\par
The contribution from Fig.2e  contains the double log $\ln^2 k_1^-$, i.e., $\ln^2\bar y$.
This double log can be obtained by making the approximation for the contribution
in the limit of $k \sim \delta$ and $k_1\sim \delta$ with $\delta \to 0$, where we denote
the momentum of the gluon 2 as $k$. Under the approximation we obtain:
\begin{eqnarray}
 F_\pi (Q) \vert_{2e} & \approx & -\frac{g_s^4}{K^-}  \frac{f^{abc}} { (p_1-k_1)^2}
     \left [ \bar u(k_2) \gamma^\alpha T^c  T^b u(p_2)
        \bar v(p_1) \gamma^\alpha T^a v(k_1) \right ]
\nonumber\\
 && \cdot \int \frac{d^4 k}{(2\pi)^4} \frac{2 p_1^+}{P^+} \frac{1}{(k^2 +i\varepsilon) ((k_1-p_1-k)^2 +i\varepsilon)
  ( k^+ +i\varepsilon)} + \cdots
\nonumber\\
     &=&  F_\pi^{(0)} (Q)  \left [ \frac{\alpha_s N_c}{4\pi}
 \left (  \ln \bar y \left ( -\frac{2}{\epsilon} +\gamma -\ln 4\pi \right )
    + \ln \bar y \ln\frac{ \bar x Q^2}{\mu^2} + \ln^2 \bar y- \frac{1}{2} \ln^2 \bar y \right ) \right ] +\cdots.
\end{eqnarray}
Again, the first double log of $\bar y$ is Type I double log, and the term $-(\ln^2 \bar y)/2$
is Type II.
Analyzing  the contribution from the region of $p_1\sim k \sim \delta$ and
the cases with the soft gluon as the gluon 1, we do not
find any other double log's.
\par
A complicated case is the contribution from Fig.2f.
We first consider the case where the gluon 2 with the momentum $k$ is soft.
The contribution in the region of  $p_1 \sim k \sim \delta$ we have the approximation:
\begin{eqnarray}
 F_\pi (Q) \vert_{2f}
   &\approx & - i \frac{g_s^4}{2 N_c K^-}
\left [ \bar u(k_2)   \gamma^\beta T^a  u(p_2)
 \bar v(p_1) \gamma_\beta T^a  v(k_1) \right ]
\nonumber\\
 && \cdot \frac{ 2 k_2^-}{P\cdot k_1} \int \frac{ d^4 k}{(2\pi)^4}
\frac{ 1 }{ (k^2 +i\varepsilon)((k-k_2)^2 +i\varepsilon)((k-p_1)^2+i\varepsilon) }+\cdots
\nonumber\\
  &=& F_\pi^{(0)} (Q) \left [ - \frac{\alpha_s}{4\pi N_c}  \left ( \left (\frac{2}{\epsilon}\right )^2
          +\frac{2}{\epsilon} \left ( -\gamma + \ln\frac{4\pi \mu^2}{\bar x y Q^2} \right )
             +\frac{1}{2} \ln^2 \frac{ 4\pi\mu^2}{\bar x y Q^2 } \right )\right ] +\cdots.
\end{eqnarray}
This contribution generates
the double log of $\ln^2 (p_1^+)$, it also generates a double log of $\ln^2 (k_2^-)$. The later
is irrelevant here for our purpose. From the above one can see that the double pole and
the associated double log $\ln^2\bar x$, i.e., $\ln^2 (p_1^+)$,  is canceled by those from Fig.2a, as discussed before.
The contribution from Fig.2f also contains a double log of $k_1^-$.
To analyzing the contribution from the region of $k\sim k_1\sim\delta$ one can use the trick in \cite{FGOC}
to decompose the production of five propagators into a sum of products of four propagators.
Then one can find the dominant contribution from the region $k\sim k_1\sim\delta$ as:
\begin{eqnarray}
 F_\pi (Q) \vert_{2f} &\approx &
     i \frac{g_s^4 k_2^- }{N_c K^- p_1\cdot k_1 }
\left [ \bar u(k_2)   \gamma^\beta T^a  u(p_2)
 \bar v(p_1) \gamma_\beta T^a  v(k_1) \right ]
\nonumber\\
   && \cdot \int \frac{ d^4 k}{(2\pi)^4}
    \frac{1 }
    {(k^2+i\varepsilon) ( (k-k_2)^2 +i\varepsilon ) ((P-k_1-k)^2+i\varepsilon)}
    +\cdots
\nonumber\\
    &=& F_\pi^{(0)} (Q)\left [ - \frac{\alpha_s}{4\pi N_c} \left ( \ln \bar y \left ( -\frac{2}{\epsilon}
        + \gamma + \ln\frac{Q^2}{4\pi \mu^2} \right ) + \ln^2 \bar y - \frac{1}{2} \ln^2 \bar y \right) \right ] +\cdots.
\end{eqnarray}
Again, the first double log of $\bar y$ is Type I double log, and
the term $-(\ln^2 \bar y)/2$. For the case that the gluon 1 is soft
with $\mu=-$ one does not find the double log of $\ln^2 k_1^-$.
Hence, with $\mu=-$ only the soft gluon 2 generates the wanted
double log's. The above results of Fig.2f are in agreement with
those in \cite{FGOC,BNP}.
\par
Performing a similar analysis of contribution from Fig.2d, we find
all relevant double log's from the region where the gluon 1 is soft.
We denote $k$ as the momentum of the gluon 1.
In the region $k_1\sim k\sim \delta$ we have the approximation:
\begin{eqnarray}
F_\pi (Q) \vert_{2d} & \approx & -i g_s^4 \frac{ 2x }{N_c x Q^2}
  \left [ \bar u(k_2) \gamma^\alpha T^a  u(p_2) \bar v(p_1) \gamma_\alpha T^a v(k_1) \right ]
\nonumber\\
   && \cdot \int \frac{d^4 k}{(2\pi)^4}
   \frac{1}{(k^2+i\varepsilon)((k+k_1)^2+i\varepsilon)((p_2+k)^2+i\varepsilon)}
\nonumber\\
   & \approx & F_\pi^{(0)}(Q) \left [ -\frac{\alpha_s}{4\pi N_c} \left ( \left (\frac{2}{\epsilon} \right )^2
     +\frac{2}{\epsilon} \left ( -\gamma +\ln \frac{ 4\pi \mu^2}{x \bar y Q^2} \right )
       +\frac{1}{2} \ln^2 \frac{4\pi \mu^2}{x \bar y Q^2} \right ) \right ] +\cdots .
\end{eqnarray}
Comparing the result in Eq.(12) for Fig.2a one realizes that the double log $\ln^2 \bar y$ in the above
is canceled by that in Eq.(12). The contributions from Fig.2c does not have
relevant double log's. Adding the soft contributions from Fig.2a, Fig.2c, Fig.2d and Fig.2f.,
all double poles are canceled.
For the region $p_1\sim k \sim \delta$ we similarly have:
\begin{eqnarray}
F_\pi (Q) \vert_{2d}  & \approx &
i g_s^4 \frac{ 4 p_2^+ k_1^-}{N_c Q^2 \bar x \bar y Q^2}
  \left [ \bar u(k_2) \gamma^\alpha T^a  u(p_2) \bar v(p_1) \gamma_\alpha T^a v(k_1) \right ]
\nonumber\\
   && \cdot
    \int \frac{d^4 k}{(2\pi)^4}
   \frac{1 }{(k^2+i\varepsilon)
   ((p_2+k)^2+i\varepsilon)( ( p_1-k_1 -k)^2 +i\varepsilon)} + \cdots
\nonumber\\
  & \approx & F_\pi^{(0)} (Q) \left [ -\frac{\alpha_s}{4 \pi N_c} \left ( \ln \bar x \left (
     -\frac{2}{\epsilon} +\gamma + \ln \frac{\bar y Q^2}{4\pi \mu^2} \right )
     + \ln^2 \bar x - \frac{1}{2} \ln^2 \bar x \right ) \right ] +\cdots.
\end{eqnarray}
Again, the first double log of $\bar x$ is Type I double log, and
the term $-(\ln^2 \bar x)/2$.
\par
From the above analysis,
one can find that the final contributions of double log's to the perturbative coefficient
$\tilde H$
only come from
Type I and Type II double log's in Fig.2b, Fig.2e, Fig.2f and Fig.2d.
Adding every double log's together we have:
\begin{eqnarray}
 {\mathcal H}(x,y,Q,\mu ) &=& 1 + \frac{\alpha_s (N_c^2-1)}{4\pi N_c} \left [ (\ln^2 \bar x + \ln^2 \bar y) -
 \frac{1}{2} (\ln^2 \bar x + \ln^2 \bar y)\right ]
\nonumber\\
   && + \frac{2\alpha_s}{3\pi} \ln\bar x \ln \bar y + \frac{\alpha_s}{4\pi} \left [ \beta_0 \ln\frac{\mu^2}{Q^2}
       -\frac{8}{3} ( 3 + \ln\bar x + \ln \bar y ) \ln \frac{\mu^2}{Q^2} \right ] +\cdots,
\end{eqnarray}
where $\cdots$ stand for non-log, i.e., rational terms. In the first line the double log's given in
$[ \cdots ]$ come from
Fig.2b, Fig.2e, Fig.2f and Fig.2d. The Type I double log's are given in the first $(\cdots)$,
the Type II double log's are given in the second $(\cdots)$.
The results in the second line are from previous results in \cite{FGOC,DR,KR,BT,BNP}.
With the above results
the double log $\ln^2 \bar x$ from Fig.2b and Fig.2d
in Eq.(19) and the double log $\ln^2\bar y$ from Fig.2e and Fig.2f in Eq.(17) need to be resummed.
\par\vskip20pt

\noindent
{\bf 4. Factorization of the Double Logs}
\par\vskip15pt\noindent
{\bf 4.1 Factorization of Type I double log's}
\par
We have seen in the last section that the origin of Type I double log's depends
on how the collinear divergence is regularized.
As discussed in detail in \cite{FMW}, this type of double log's can be factorized
by introducing NLCWF. The definition of NLCWF is:
\begin{eqnarray}
\tilde  \phi(x, \zeta_{\tilde u}, \mu) & = & \ \int \frac{ d z^- }{2\pi}
    e^{ik^+z^- }
 \frac{ \langle 0 \vert \bar q(0) L_{\tilde u}^\dagger (\infty, 0)
  \gamma^+ \gamma_5 L_u (\infty,z^-n ) q(z^-n ) \vert \pi^0(P) \rangle}
  { \tilde S( z^-, \zeta_u) },
\nonumber\\
  L_{\tilde u} (\infty, z) &=& P \exp \left ( -i g_s \int_{0} ^{\infty} d\lambda
     \tilde u\cdot G (\lambda \tilde u+ z ) \right ),  \ \ \  \tilde u^\mu =( \tilde u^+, \tilde u^-,0,0) .
\end{eqnarray}
In the above the limit $\tilde u^- \gg \tilde u^-$ should be taken.
The main difference between LCWF and NLCWF is the gauge link in definitions. In LCWF the gauge link
is obtained from $L_{\tilde u}$ by setting $\tilde u^+ =0$. It should be noted
that the limit $\tilde u^+ \to 0$ or $\tilde u^- \gg \tilde u^+$ is
taken after integrations of loop-momenta in the definition\cite{CS,CSS,JMY,JMYG}.
Otherwise the above definition is reduced to that of LCWF.
The defined NLCWF depends on an extra parameter
\begin{equation}
\zeta^2_{\tilde u} = \frac{2 \tilde u^- (P^+)^2}{\tilde u^+}\approx \frac{ 4 (\tilde u\cdot P)^2}{\tilde u^2}.
\end{equation}
The factor $ \tilde S( z^-, \zeta_{\tilde u})$ is the vacuum expectation value of a product of four gauge links.
Details can be found in \cite{FMW}.
In \cite{FMW} it has been shown that
there is an interesting relation between LCWF and NLCWF. It reads:
\begin{equation}
\tilde \phi (x,\zeta,\mu) = \int_0^1 dy C(x,y,\zeta,\mu) \phi (y,\mu),
\end{equation}
where the function $C$ can be calculated perturbatively. The relation is in fact a factorization
relation. NLCWF contains the same collinear singularities as LCWF. Therefore the function $C$
does not contain any soft divergence.
Detailed results of $C$ at one-loop level can be found in \cite{FMW}.
For our purpose to do the resummation at leading order, it is enough to consider the function:
\begin{equation}
\frac{ \hat C (x,\zeta,\mu )}{\bar x} =\int_0^1 \frac{dy }{\bar y} C(y,x,\zeta,\mu).
\end{equation}
From \cite{FMW} we have:
\begin{eqnarray}
\hat C (x,\zeta,\mu )
   &=& 1 -\frac{\alpha_s(\mu) (N_c^2-1)}{4\pi N_c} \left \{ \frac{1} {x}
\left [-\ln^2 \bar x + \ln \bar x \ln\frac{\mu^2}{\zeta^2 } + 2 {\rm Li}_2 (\bar x)  -\frac{\pi^2}{3} \right ]
\right.
\nonumber\\
   &&  \left.
    + \frac{1}{2} \ln\frac{\mu^2}{\zeta^2 \bar  x^2}
 +  \frac{1}{2}\ln\frac{\mu^2}{\zeta^2 x^2}
        + \frac{\pi^2}{3} +2  \right \} + {\mathcal O} (\alpha_s^2).
\end{eqnarray}
For $\pi$ in the final state moving with the momentum $K^\mu =(0,K^-,0,0)$ one can introduce the NLCWF
$\tilde  \phi(x, \zeta_{\tilde v}, \mu)$
with the replacement $\tilde u \to \tilde v$ with $\tilde v^\mu = (\tilde v^+,\tilde v^-,0,0)$ and
$\tilde v^+ \gg \tilde v^-$. The parameter
$\zeta_{\tilde v}$ is defined as:
\begin{equation}
\zeta^2_{\tilde v} = \frac{2 \tilde v^+ (K^-)^2}{\tilde v^- }\approx \frac{ 4 (\tilde v\cdot K)^2}{\tilde v^2}.
\end{equation}
\par
Because NLCWF and LCWF contain the same collinear singularities, one can also use NLCWF
to factorize the form factor.
With these NLCWF's the form factor can be factorized as:
\begin{equation}
F_\pi (Q) \approx \frac{8 \pi \alpha_s}{9 Q^2}
\int_0^1 \frac {dx}{\bar x} \frac{dy}{\bar y}
\tilde \phi (x,\zeta_{\tilde u}, \mu) \tilde \phi  (y,\zeta_{\tilde v}, \mu)
\hat {\mathcal H}(x,y,\zeta_{\tilde u},\zeta_{\tilde v},Q,\mu),
\end{equation}
Comparing two factorizations we have the relation:
\begin{equation}
\hat{\mathcal H}(x,y,\zeta_{\tilde u},\zeta_{\tilde v},Q,\mu) =\hat C^{(-1)} (x,\zeta_{\tilde u},\mu )
\hat C^{(-1)} (y,\zeta_{\tilde v},\mu )
  {\mathcal H}(x,y,Q,\mu),
\end{equation}
at one-loop level. At higher orders, the relation can be a convolution with the functions $C's$.
We find then
\begin{eqnarray}
\hat {\mathcal H}(x,y,\zeta_{\tilde u},\zeta_{\tilde v},Q,\mu) &=&  1 + \frac{\alpha_s (N_c^2-1)}{4\pi N_c}
\biggr [  \frac{\bar x}{x}\ln^2 \bar x + \frac{\bar y}{y}\ln^2 \bar y -
 \frac{1}{2} (\ln^2 \bar x + \ln^2 \bar y)
\nonumber\\
 &&  + \frac{1}{ x}\ln\bar x \ln\frac{\mu^2}{\zeta_{\tilde u}^2} +
  \frac{1}{ y}\ln\bar y \ln\frac{\mu^2}{\zeta_{\tilde v}^2} +\ln \frac{\mu^2}{\zeta_{\tilde u}^2}+
  \ln \frac{\mu^2}{\zeta_{\tilde v}^2}  \biggr ]
\nonumber\\
    && + \frac{2\alpha_s}{3\pi} \ln\bar x \ln \bar y + \frac{\alpha_s}{4\pi} \left [ \beta_0 \ln\frac{\mu^2}{Q^2}
       -\frac{8}{3} ( 3 + \ln\bar x + \ln \bar y ) \ln \frac{\mu^2}{Q^2} \right ] +\cdots,
\end{eqnarray}
where we only give these terms explicitly: the double log terms and the scale-dependent terms.
For $\bar x \to 0$ or $\bar y \to 0$ we find that in $\hat{\mathcal H}$ there are no Type I double log's.
Therefore, the Type I double log's are factorized.
\par\vskip15pt
\noindent
{\bf 4.2 Factorization of Type II double log's}
\par
In this section we will introduce soft factors to factorize Type II double log's. We first discuss
how to deal the double log from Fig.2b. As discussed before, in Eq.(13,14) the term $-1/2 \ln^2 \bar x$
which is divergent with $\bar x \to 0$ is generated from the region in which the gluon 1 is soft.
The contribution from the soft gluon can be subtracted by using non-light-like
gauge link\cite{JH}. The subtraction can be done by replacing
the quark propagator attached by the gluon 1 with the eikonal propagator:
\begin{equation}
\frac{1}{(p_2-k)^2 -m^2 +i\varepsilon} \approx \frac{1}{-2p_2^+ k^- +i\varepsilon}
 \to \frac{1}{- v\cdot k + i\varepsilon},
\end{equation}
with $v^\mu =(v^+,v^-,0,0)$ and $v^+ \gg v^-$. Combining factors from numerators one can identify
the soft contribution as the contribution which is obtained from Fig.2b by replacing the quark line
attached with the soft gluon with a gauge link along the direction $v$. Then one realizes
that the soft contribution can be subtracted or factorized with the soft factor like
\begin{equation}
 \langle 0 \vert V(x,-\infty, v) G^{a,\mu} (x) G^{b,\nu} (0) \vert 0 \rangle,
\end{equation}
$V$ is the gauge link along the direction $v$ from $-\infty$ to $x$.
At tree-level the soft factor is just the free propagator of gluon.
This type of subtraction has been discussed in \cite{ASY}.
\par
The above soft factor as it stands is not gauge invariant. To factorize the Type II double log in Fig.2b
with a gauge invariant soft factor one can modify or extend the above soft factor. For this we consider
the following soft factor defined with field strength tensor $G^{\mu\nu}$ and gauge links:
\begin{eqnarray}
 S(\tilde k )  &= & \int d^4 x e^{i \tilde k\cdot x} {\rm Tr} \langle 0 \vert T \left [
    V(x,\infty, w) G^{- \mu}(x) V(x,-\infty, v)
\right.
\nonumber\\
    && \left.  \  V^\dagger (0,-\infty, \tilde w) G^{-\nu} (0) V^\dagger (0, \infty, w) \right ] \vert 0 \rangle
    \left ( g_{\mu\nu} - n_\mu l_\nu - n_\nu l_\mu \right ) ,
\end{eqnarray}
with $\tilde k^\mu = (x P^+, -K^-,0,0)$.
The gauge link along the direction $v$ pointing to future or starting from the past is defined as:
\begin{eqnarray}
V (x,\infty, v) &=& P \exp \left ( -i g_s \int_0^{\infty} d\lambda
     u\cdot G (\lambda v + x) \right ),
\nonumber\\
V(x,-\infty, v) &=& P \exp \left ( -i g_s \int^0_{-\infty} d\lambda
     u\cdot G (\lambda v + x) \right ).
\end{eqnarray}
The three vectors for directions of gauge links are defined as the following:
We take the vector $v$ as $v^\mu =(v^+,v^-,0,0)$.
The other two vectors $w$ and $\tilde w$ are transverse.
The three vectors are chosen as:
\begin{equation}
   v\cdot w =0, \ \ \ \  v\cdot \tilde w =0, \ \ \ \ w\cdot \tilde w =0.
\end{equation}
In the soft factor the limit $v^+\gg v^-$ is taken.
The reason for taking the field strength tensor $G^{-\mu}$ is because
with other components of $G^{\mu\nu}$ the soft factor under the limit
$v^+\gg v^-$ can have the divergence
as $1/v^-$.
\par

\begin{figure}[hbt]
\begin{center}
\includegraphics[width=15cm]{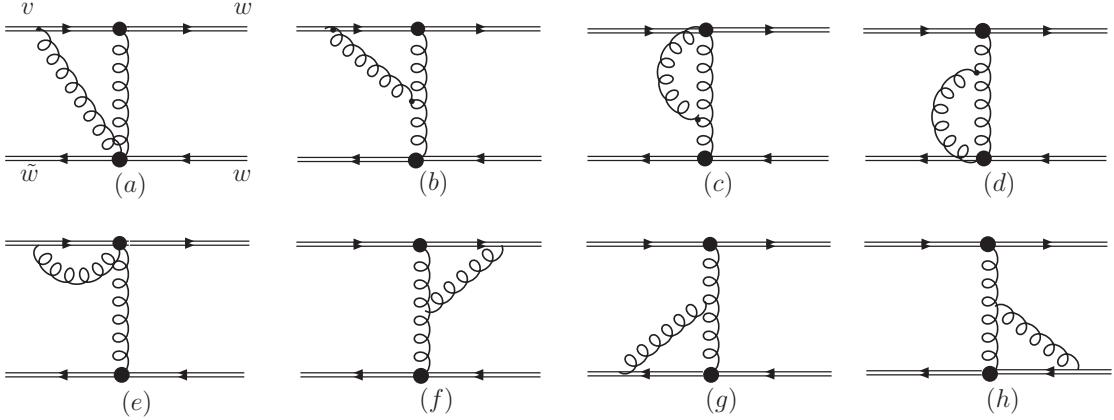}
\end{center}
\caption{Part of one-loop corrections. }
\label{Feynman-dg3}
\end{figure}
\par
At the tree-level we have:
\begin{equation}
S^{(0)} (\tilde k ) = -i (N_c^2-1) \frac{ (l\cdot \tilde k)^2}{\tilde k^2}.
\end{equation}
At one-loop level, many diagrams exist. In Fig.3 and Fig.4 diagrams for one-loop corrections
are given. Some diagrams are automatically zero because of
the directions of gauge links in Eq.(34). These diagrams are not drawn in Fig.3 and Fig.4.
We first study the contribution from Fig.3b which corresponds to the soft contribution in Fig.2b.
The correspondence is realized by replacing the quark propagator attached by the gluon 1 in Fig.2b
with the gauge link along the direction $v$.
It is straightforward to obtain the contribution:
\begin{equation}
S(q)\vert_{3b} = S^{(0)} (\tilde k ) \cdot \frac{\alpha_s N_c}{4\pi}
\left [ -\frac{1}{2} \ln^2 \frac{\bar x Q^2}{ \zeta_v^2} - \ln \frac{\bar x Q^2}{ \zeta_v^2}
+\frac{1}{2}  \ln\left ( \frac{\mu^2}{-\tilde k^2}\right )
 -\frac{2}{3} \pi^2 -\frac{1}{4} +{\mathcal O} (\zeta_v^{-2})\right ],
\label{sg}
\end{equation}
with
\begin{equation}
\zeta_v^2 = \frac{ 2 v^+ (K^-)^2}{v^-}.
\end{equation}
From the above result one can see that the contribution from Fig.3b to the soft factor
contains the same Type double log as that in the contribution from Fig.2b to the
form factor, as expected.
\par
The contribution from Fig.3e is exactly zero. The contributions from other diagrams in Fig.3 are:
\begin{eqnarray}
S(\tilde k)\vert_{3a} &=&   0 + {\mathcal O} (\zeta_v^{-2}),
\nonumber\\
S(\tilde k)\vert_{3e} &=& 0,
\nonumber\\
S(\tilde k)\vert_{3f} &=& S(\tilde k)\vert_{3g} = S(\tilde k)\vert_{3h} = S^{(0)}(\tilde k) \frac{\alpha_s N_c}{16\pi}
\left [ 2 \ln\left ( \frac{\mu^2}{-\tilde k^2}\right )  +3 \right ],
\nonumber\\
S(\tilde k)\vert_{3c} &=& S(\tilde k)\vert_{3d} =- S^{(0)}(\tilde k) \frac{3 \alpha_s N_c}{4\pi}
\left [  \ln\left ( \frac{\mu^2}{-\tilde k^2}\right )  +1 \right ].
\end{eqnarray}

\par\vskip20pt

\begin{figure}[hbt]
\begin{center}
\includegraphics[width=12cm]{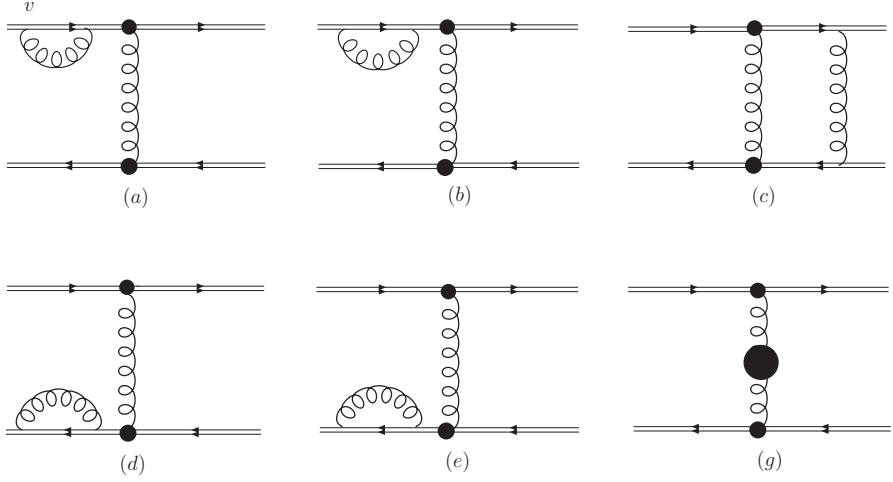}
\end{center}
\caption{Self-energy-Corrections of gauge links. }
\label{Feynman-dg3}
\end{figure}
\par
There are also one-loop corrections from the self-energy of gauge links, exchanges of gluon between
gauge links and the self-energy of gluon.  The corresponding diagrams are given
in Fig.4. In contrast to the contributions from Fig.3., which do not contain any soft divergence,
the corrections from the self-energy of gauge links, exchanges of gluon between
gauge links in Fig.4. contain I.R. singularities.
These contributions can be subtracted by
modifying the definition of $S(\tilde k)$ in Eq.(32).
For this purpose we introduce
\begin{equation}
S_0(x) = \frac{1}{N_c} {\rm Tr} \langle 0 \vert T \left [
    V(x,\infty, w)  V(x,-\infty, v)
    V^\dagger (0,-\infty, \tilde w)
    V^\dagger (0, \infty, w) \right ] \vert 0 \rangle ,
\end{equation}
and modify $S(\tilde k)$ to $\tilde S(\tilde k)$:
\begin{equation}
 \tilde S(\tilde k)  = \int d^4 x e^{i \tilde k \cdot x}  \frac{ S(x) }{S_0(x)}.
\end{equation}
With the modification, The contributions from Fig.4a to Fig.4e are subtracted, i.e.,
there is no contribution from Fig.4a to Fig.4e to $\tilde S(\tilde k)$.
The remaining contribution from Fig.4 to $\tilde S(q)$ is only from Fig.4g. It is
\begin{equation}
\tilde S(\tilde k)\vert_{4g} =  S^{(0)}(\tilde k) \frac{\alpha_s}{4\pi} \left [ N_c \left (
\frac{5}{3}  \ln\frac{\mu^2}{-q^2} + \frac{31}{9}\right ) -\frac{2}{3} N_f \left ( \ln\frac{\mu^2}{-q^2}
+\frac{5}{3} \right ) \right ].
\end{equation}
The final result for $\tilde S(q)$ at one-loop level reads:
\begin{eqnarray}
\tilde S(\tilde k)&=& S^{(0)}(\tilde k) \left [ 1 - \frac{ \alpha_s N_c}{8\pi}
\left ( \ln^2 \frac{\bar x Q^2}{\zeta_v^2}
   + 2 \ln \frac{\bar x Q^2}{\zeta_v^2} \right )
  -\frac{ \alpha_s }{48\pi} \left ( 31 N_c + 8 N_f \right )
      \ln \frac{\mu^2}{-\tilde k^2}
\right.
\nonumber\\
    && \left. \ \ \ \ \ \ \ \ \ \ \  -\frac{\alpha_s}{4\pi} \left ( N_c \left (\frac{2}{3}\pi^2 + \frac{5}{9} \right )
      + \frac{10}{9} N_f \right )     \right ].
\end{eqnarray}
With the soft factor $\tilde S$ we can factorize the Type II double log term in Fig.2b.


\begin{figure}[hbt]
\begin{center}
\includegraphics[width=12cm]{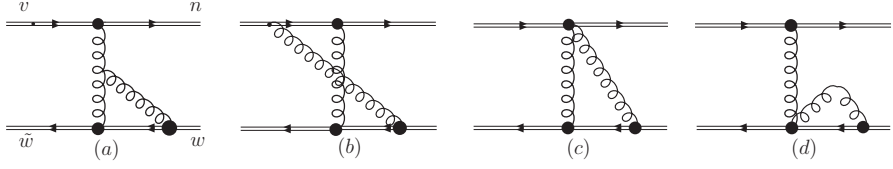}
\end{center}
\caption{The leading order contributions of the correlation function $S_D$. }
\label{Feynman-dg3}
\end{figure}
\par
Now we turn to the Type II double log $\ln^2\bar x$ in Fig. 2d.
It is rather difficult to construct a soft factor to factorize the double log. However,
at one-loop level we can build the following soft factor which can be used for the factorization
at least at one-loop level:
\begin{eqnarray}
 S_D(\tilde k) &=& \frac{2 i l\cdot \tilde k}{\tilde k^2} \int d^4 x e^{i \tilde k \cdot x} {\rm Tr} \langle 0 \vert T \left [
    V(x,\infty, n) G^{- \mu}(x) V(x,-\infty, v) V^\dagger (0,-\infty, \tilde w) G^{-\nu} (0)
\right.
\nonumber\\
    && \left. \ \ \ \ \ \ \ \ \ \ \ \ \ \ \ \ \ \ \ \
    \ \ \ \ \  \left ( D^+ V^\dagger (0, \infty, w) \right ) \right ] \vert 0 \rangle
     \left ( g_{\mu\nu} - n_\mu l_\nu - n_\nu l_\mu \right ) .
\end{eqnarray}
This correlation function consists of a covariant derivative of one
gauge link. It is gauge invariant. The leading order contribution comes from the
gluon change where the gluon is emitted by the term of the covariant
derivative of the gauge link.
In the limit of $v^- \to 0$ the leading order contributions to $S_D$ come from the diagrams given
in Fig.5. The contribution from Fig.5b will contain the double log corresponding to that from Fig.2d.
The contribution from Fig.5d is
zero in the dimensional regularization.
The contributions from each diagrams in Fig.5. are:
\begin{eqnarray}
S_D(\tilde k ) \vert_{5a}
 &=& S^{(0)}(\tilde k) \frac{\alpha_s N_c }{8 \pi}  \left [  \ln \left ( \frac{\mu^2}{-\tilde k^2} \right ) +\frac{13}{6}
  \right ],
\nonumber\\
S_D(\tilde k) \vert_{5b}
     &=& S^{(0)}(\tilde k) \frac{\alpha_s}{8\pi N_c} \left [ \ln^2 \frac{\bar x Q^2}{\zeta^2_v}
          +2 \ln \frac{\bar x Q^2}{\zeta^2_v} -\ln\left ( \frac{\mu^2}{-\tilde k^2}\right )
           +\frac{4}{3}\pi^2 -1 + {\mathcal O}(\zeta_v^{-2})
             \right ],
\nonumber\\
S_D(\tilde k )\vert_{5c} &=& S^{(0)}(\tilde k) \frac{\alpha_s N_c }{8\pi}  \left [ \ln \left (\frac{\mu^2}{-\tilde k^2} \right )
   +1  \right ],
\nonumber\\
S_D(\tilde k) \vert_{5d} &=& 0.
\end{eqnarray}
From the above result one can see that the double log $\ln^2\bar x$ in the contribution
from Fig.2d. to the form factor is correctly produced with the contribution from Fig.5b. to the soft factor.
To factorize all Type II double log term of $\bar x$, we can now introduce the following
soft factor combining above results:
\begin{eqnarray}
S_v(\bar x Q^2,\zeta_v, \mu) &=&  \frac{1} {S^{0}(\tilde k )} \left  (\tilde S(\tilde k) + S_D(\tilde k) \right )
\nonumber\\
   &=&  1 - \frac{ \alpha_s ( N_c^2-1)}{8\pi N_c } \left ( \ln^2 \frac{\bar x Q^2}{\zeta_v^2}
   + 2 \ln \frac{\bar x Q^2}{\zeta_v^2} \right )
 -\frac{ \alpha_s }{48\pi} \left ( 19 N_c + 8 N_f +\frac{6}{N_c}  \right )
      \ln \frac{\mu^2}{ \bar x Q^2}
\nonumber\\
     &&   +\cdots +{\mathcal O }(\alpha_s^2)+{\mathcal O}(\zeta_v^{-2}),
\end{eqnarray}
where $\cdots$ denote constant terms. This soft factor will be used to re-factorize
the double log related to $q^+$, i.e., $\ln^2 x$ with $q$ given in Eq.(10).
\par
The Type double log of $\bar y$ in the contributions from Fig.2e and Fig.2f can be handle in a similar way
by introducing another soft factor. The soft factor can be obtained
from $S_v$ through the time-reversal transformation, where we replace the transformed vector $v$
with $u$ and the transformed momentum $\tilde k$ with $(P^+, -\bar y K^-,0,0)$.
The vector $u$ is given by $u^\mu =(u^+,u^-,0,0)$ with $u^-\gg u^+$. We denote
the obtained soft factor as $S_u(\bar y Q^2, \zeta_u, \mu)$.
Its one-loop result can be read from $S_v$ as:
\begin{eqnarray}
S_u(\bar y Q^2,\zeta_u, \mu) &=&  1 - \frac{ \alpha_s ( N_c^2-1)}{8\pi N_c } \left ( \ln^2 \frac{\bar y Q^2}{\zeta_u^2}
   + 2 \ln \frac{\bar y Q^2}{\zeta_u^2} \right )
 -\frac{ \alpha_s }{48\pi} \left ( 19 N_c + 8 N_f +\frac{6}{N_c}  \right )
      \ln \frac{\mu^2}{\bar y Q^2}
\nonumber\\
     &&   +\cdots +{\mathcal O }(\alpha_s^2)+{\mathcal O}(\zeta_u^{-2}),
\nonumber\\
   \zeta_u^2 &=& \frac{ 2u^- (P^+)^2}{u^+}.
\end{eqnarray}
\par
With the introduced two soft factors we can now factorize the Type II double log's by writing
the factorized form of the form factor as:
\begin{eqnarray}
F_\pi (Q^2) &\approx & \frac{8 \pi \alpha_s}{9 Q^2}
\int_0^1 \frac {dx}{\bar x} \frac{dy}{\bar y}
\tilde \phi (x,\zeta_{\tilde u}, \mu) \tilde \phi  (y,\zeta_{\tilde v}, \mu)
S_u(\bar x Q^2,\zeta_u, \mu) S_v(\bar y Q^2,\zeta_v, \mu)
\nonumber\\
    &&   \ \ \ \  \cdot \tilde {\mathcal H} (x,y,\zeta_{\tilde u},\zeta_{\tilde v},\zeta_u,\zeta_v,Q,\mu),
\end{eqnarray}
with
\begin{eqnarray}
{\mathcal H}(x,y,\zeta_{\tilde u},\zeta_{\tilde v},\zeta_u,\zeta_v,Q,\mu) &=&  1 + \frac{2 \alpha_s }{3\pi }
\left [  \frac{\bar x}{x}\ln^2 \bar x + \frac{\bar y}{y}\ln^2 \bar y + \ln\bar x \ln\frac{Q^2}{\zeta_v^2}
   +\frac{1}{2}\ln^2\frac{Q^2}{\zeta_v^2} + \ln\frac{\bar x Q^2}{\zeta_v^2}
\right.
\nonumber\\
 && \left.  +\ln\bar y \ln\frac{Q^2}{\zeta_u^2}
   +\frac{1}{2}\ln^2\frac{Q^2}{\zeta_u^2} + \ln\frac{\bar y Q^2}{\zeta_u^2}
   + \frac{1}{ x}\ln\bar x \ln\frac{\mu^2}{\zeta_{\tilde u}^2} +
  \frac{1}{ y}\ln\bar y \ln\frac{\mu^2}{\zeta_{\tilde v}^2}
\right.
\nonumber\\
  && \left.  +\ln \frac{\mu^2}{\zeta_{\tilde u}^2}+
  \ln \frac{\mu^2}{\zeta_{\tilde v}^2}  \right ] + \frac{ 83 \alpha_s}{24\pi} \ln \frac{\mu^2}{Q^2}
+ \frac{2\alpha_s}{3\pi} \ln\bar x \ln \bar y
\nonumber\\
 && + \frac{\alpha_s}{4\pi} \left [ \beta_0 \ln\frac{\mu^2}{Q^2}
       -\frac{8}{3} ( 3 + \ln\bar x + \ln \bar y ) \ln \frac{\mu^2}{Q^2} \right ] +\cdots,
\end{eqnarray}
where we have taken $N_c=3$ and $N_f=3$. It is clear that the above expression does not contain
the divergent double log $\ln^2 \bar x$ or $\ln^2 \bar y$ for $\bar x\to 0$ or $\bar y\to 0$, respectively.
All these divergent double log's are contained in the soft factors and NLCWF's.
\par\vskip20pt
\noindent
{\bf 5. Resummation and Numerical Result}
\par\vskip15pt
\par
To re-sum those double log's in the original $H$ we express the form factor
in LCWF's through functions $\hat C$'s and take all scales at $Q^2$:
\begin{equation}
F_\pi (Q) \approx \frac{8 \pi \alpha_s}{9 Q^2}
\int_0^1 \frac {dx}{\bar x} \frac{dy}{\bar y}
 \phi (x, Q)  \phi  (y, Q) \hat C(x,Q,Q) \hat C( y,Q,Q)
S_v(\bar x Q^2,Q, Q) S_u(\bar y Q^2,Q, Q) {\mathcal H}_{r} (x,y),
\end{equation}
with the simple hard part:
\begin{eqnarray}
   {\mathcal H}_r (x,y) &=& {\mathcal H}(x,y,Q,Q,Q,Q,Q,Q)
\nonumber\\
        &=& 1 + \frac{2 \alpha_s }{3\pi }
\left [  \frac{\bar x}{x}\ln^2 \bar x + \frac{\bar y}{y}\ln^2 \bar y + \ln\bar x + \ln\bar y
+  \ln\bar x \ln \bar y \right ] +\cdots,
\end{eqnarray}
so that ${\mathcal H}_r (x,y)$ does not contain those double log's.
\par
The resummation can be done as in the following. For $\hat C$-function we can follow
\cite{FMW} to express $\hat C(x,Q,Q)$ in term of $\hat C(x,\mu_0/\sqrt{\bar x}, \mu_0)$ with
$\mu_0$ determined by:
\begin{equation}
    \alpha_s (\mu_0) = \bar x \alpha_s(Q).
\end{equation}
It should be noted that for a given large scale $Q$ hence a small $\alpha_s(Q)$ $\alpha_s(\mu_0)$
is always smaller than  $\alpha_s(Q)$ because the asymptotic freedom of QCD. Hence
an pertrubative expansion of the $\hat C$ function in $\alpha_s(\mu_0)$
still make sense.
We have:
\begin{eqnarray}
\hat C(x,Q,Q) &=& \exp \left \{ - \frac{8}{3\beta_0}\left [ \ln \bar x
  -\frac{\beta_0}{4\pi} \bar x \alpha_s(Q) \ln \bar x -\bar x  +1 \right]
  \left ( \frac{\ln \bar x}{ x} +1 \right ) \right \}
\nonumber\\
    && \ \cdot \hat C(\bar x,  \mu_0/\sqrt{\bar x},\mu_0),
\nonumber\\
\hat C(x,  \mu_0/\sqrt{\bar x},\mu_0) &=& 1 -\frac{\alpha_s (\mu_0)}{3\pi} \left [ \frac{1}{ x}
\left ( 2 {\rm Li}_2 (\bar x) -\frac{\pi^2}{3} \right ) -\frac{1}{2}\ln x +\frac{\pi^2}{3} +2 \right ]
   + {\mathcal O}(\alpha_s^2).
\end{eqnarray}
All double log's are now re-summed in the exponential factor.
\par
For the resummation of the soft factor $S_v$ we have the following $\zeta$-evolution equation:
\begin{equation}
\frac{\partial S_v(\bar x Q^2, \zeta_v,\mu)}{\partial \ln\zeta_v^2}
  = - \frac{2\alpha_s}{3\pi} \left ( \ln \frac{\zeta_v^2}{\bar x Q^2} -1 \right ) S_v(\bar x Q^2, \zeta_v, \mu).
\end{equation}
Using this equation we can express
$S_v(\bar x Q^2,Q, Q)$ in term of $S_v(\bar x Q^2 \sqrt{\bar x} Q, Q)$:
\begin{eqnarray}
S_v(\bar x Q^2,Q, Q) &=& \exp\left[ -\frac{\alpha_s(Q)}{3\pi} \left ( \ln^2\bar x +2 \ln \bar x \right ) \right ]
S_v(\bar x Q^2 \sqrt{\bar x} Q, Q),
\nonumber\\
S_v(\bar x Q^2, \sqrt{\bar x} Q, Q) &=&
 1 -  \alpha_s(Q) ( \cdots )  +{\mathcal O }(\alpha_s^2),
\end{eqnarray}
where $(\cdots)$ contains only constant terms, i.e., no log's.
Doing the same for $S_u$
we finally have the resummed form factor:
\begin{eqnarray}
F_\pi (Q) &=& \frac{ 8\pi \alpha_s(Q)}{9}
\int_0^1 \frac {dx}{\bar x} \frac{dy}{\bar y}
 \phi (x, Q)  \phi  (y, Q)  e^{-S(\bar x,Q) - S(\bar y,Q) },
\nonumber\\
S(\bar x,Q) &=& \frac{8}{3\beta_0}\left [ \ln \bar x
  -\frac{\beta_0}{4\pi} \bar x \alpha_s(Q) \ln \bar x -\bar x  +1 \right]
  \left ( \frac{\ln \bar x}{ x} +1 \right ) +\frac{\alpha_s(Q)}{3\pi} \left ( \ln^2\bar x + 2\ln \bar x \right ),
\end{eqnarray}
since we work at leading order re-summation, we have to take ${\mathcal H}_r$ at tree-level, i.e.,
${\mathcal H}_r=1$. The above formula
is applicable when one knows LCWF's at the large scale $Q$. If LCWF's are determined
in a lower scale $\mu$, one can use evolution equation to express
$\phi(x,Q)$ in $\phi(x,\mu)$. In this case we have:
\begin{eqnarray}
F_\pi (Q) &=& \frac{ 8\pi \alpha_s(Q)}{9}
\int_0^1 \frac {dx}{\bar x} \frac{dy}{\bar y}
 \phi (x, \mu)  \phi  (y, \mu)  e^{-S(\bar x,Q) - S(\bar y,Q) + {\mathcal K}(x,y,Q,\mu) },
\nonumber\\
{\mathcal K}(x,y,Q,\mu) &=& -\frac{8}{3\beta_0} \ln\frac{\alpha_s(Q)}{\alpha_s(\mu)} \left (
   3 +\ln (\bar x\bar y) \right ).
\end{eqnarray}
\par


\begin{figure}[hbt]
\begin{center}
\includegraphics[width=12cm]{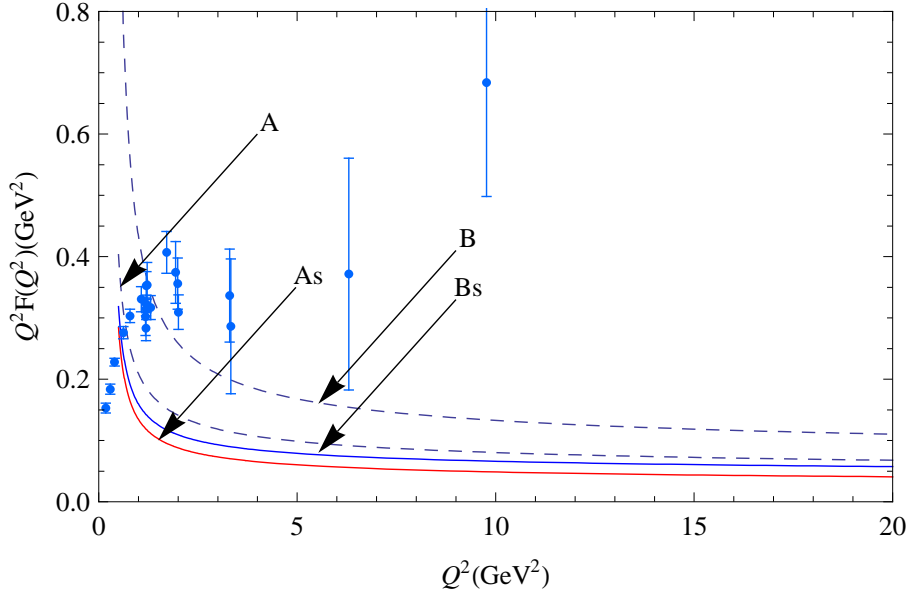}
\end{center}
\caption{Numerical results for the form factor. Curve-$A$ and Cure-$A_S$ are drawn by using LCWF
in Eq.(57) without and with the resummation. Curve-$B$ and Cure-$B_S$ are drawn by using LCWF
in Eq.(58) without and with the resummation. }
\label{Feynman-dg6}
\end{figure}
\par
We will use our resummation formula in Eq.(55) and Eq.(56) to give our numerical results.
In our formulas the nonperturbative input is the LCWF. The LCWF has the asymptotic form
if $\mu$ goes to $\infty$:
\begin{equation}
  \phi (x, \mu) = 6 x(1-x) f_\pi + \cdots,
\end{equation}
where $\cdots$ stand for terms which are zero in the limit $\mu \to \infty$.
The LCWF can be expanded with Gegenbauer
polynomials\cite{BL}.
A model for $\phi$ has been proposed by truncating the expansion\cite{BrFi}:
\begin{equation}
\phi (x, \mu) = 6 f_\pi x(1-x ) \left ( 1
+  \phi_2 (\mu) C_2^{3/2}(2x-1) \right ) ,
\end{equation}
where $\phi_2 (\mu_0)$ is determined by QCD sum-rule method at $\mu_0 =1$GeV\cite{BrFi}:
\begin{equation}
\phi_2 (\mu_0=1{\rm GeV}) =0.44.
\end{equation}
We will use these two types of LCWF to give our numerical results.
We will use Eq.(55) with the asymptotic form of $\phi$ to make our numerical
predictions. For LCWF given in Eq.(58) we use Eq.(56). We take the $\Lambda$-parameter
as $\Lambda=237$MeV. The numerical results are represented in Fig.6..
Our numerical results do not strongly depend on the value of $\Lambda$.
There is only a little change if we change $\Lambda$ from $100$MeV to $300$MeV.
\par
Our numerical results show that the resummation has significant effects. For the asymptotic
LCWF the difference between the resummed- and unresummed form factor can be from about $30\%$.
The difference with LCWF in Eq.(58) is about $50\%$. We notice that the resummed form factor is in general
smaller than the unresummed. In \cite{BNP} a detailed study for numerical predictions
at one-loop level has been done. With various models of LCWF's it has been shown
that the one-loop correction is in general positive. The correction is large and can be at order from $40\%$
to $100\%$. This brings up the question if the perturbaive expansion
is reliable. Because our resummed form factor becomes smaller, we expect that the situation
can be improved, or at least partly improved. But this needs a detailed study by including higher
order corrections.
\par
In Fig.6 we also give the experimental result
from \cite{Exp}. In the case of $\gamma^* \pi \to \gamma$ we have found that the experimental data in the region
$3{\rm GeV}^2 < Q^2 < 10{\rm GeV}^2$ can be described fairly
well with the resummed form factor\cite{FMW}.
But in the case here, we fail to re-produce the experimental results.
The similar situation also appears in the study at one-loop level in \cite{BNP}, where
the experimental results can not be reproduced with various models of LCWF.
This problem deserves a further study.

\par\vskip20pt
\noindent
{\bf 6. Conclusion}
\par\vskip15pt\noindent
\par
The hard part at one-loop level in the collinear factorization
for the form factor in $\gamma^* \pi \to \pi$
contains double log terms of $\bar x$ and $\bar y$. At $n$-loop level it is expected
that the large log terms like $\ln^{2n}\bar x$ or $\ln^{2n}\bar y$ appear.
These double log's are dangerous and  can result in that
the perturbative
expansion of the hard part becomes a divergent expansion.
A resummation of these terms with
a simple exponentiation can not be done because it results in divergent
results.
In this work we have studied the resummation of these double log terms.
For this purpose we have identified the origin of the double log's in detail
in the first step. In the second step we have employed
the concept of factorization.
We have re-factorized the form factor so that the
hard part does not contain double log's. In the re-factorization
NLCWF instead of LCWF and soft factors are introduced
to capture these double log terms.
\par
Non light-like gauge links are used in NLCWF and soft factors.
These links play important role in the re-factorization and resummation. Because
the gauge links are not light-like,
the introduced NLCWF and soft factors contain extra scales beside the renormalization scale $\mu$.
Using evolutions of these scales we are able to resum the double log terms
into two exponential factors, one is for $\ln^2\bar x$, while another is
for $\ln^2\bar y$.
An interesting aspect of our approach for the resummation
is that the resummed form factor only contains LCWF's as nonperturbative quantities.
Every ingredient in the resummation except LCWF's can be calculated perturbatively.
\par
With the resummation we have given numerical results for the form factor
with two choices of LCWF. We have found that the resummation
has significant effects. Between the resummed- the unresummed form factor
the numerical difference  is at level about $30\% -50\%$.
With various models of LCWF's it has been shown in \cite{BNP}
that the one-loop correction is in general positive. The correction is large and can be at order from $40\%$
to $100\%$. Because of the large correction, the perturbative prediction
can be unreliable. With the structure of the exponentials for the resummation
one can find that the resummed form factor becomes smaller than the unresummed.
With this fact we expect that the situation with the perturbative expansion
can be improved, or at least partly improved. But this needs a detailed study by including higher
order corrections.
\par
In the case of $\gamma^* \pi \to \gamma$ we have found that the experimental data in the region
$3{\rm GeV}^2 < Q^2 < 10{\rm GeV}^2$ can be described fairly
well with the resummed form factor\cite{FMW}. It is frustrated in the case of $\gamma^*\pi \to \pi$
if one compares theoretical results of collinear factorization with  experimental data.
With or without resummation, the experimental data at higher $Q^2$
is not in agreement with perturbative results. This deserves a further study
in experiment and in theory.
In this work we have performed
the resummation at one-loop level.
It is possible to extend our work to the resummation
of the remaining single log terms and beyond one-loop level.

\par\vskip20pt
\noindent
{\bf Appendix: Gauge Variance in the $k_T$-Factorization}
\par\vskip15pt
\par
In the collinear factorization the transverse momenta of partons entering hard scattering are
neglected at leading twist. As we have seen that the perturbative part in Eq.(9) contains double log's at higher order
and is divergent when the momentum fraction $\bar x$ or $\bar y$ approaches zero.
It has been suggested that
for small longitudinal momenta one may need to take the transverse momenta
into account. This leads to
the so-called $k_T$-factorization. The $k_T$-factorization has been widely used
in exclusive $B$-decays. Although it has been widely used, the factorization has not been
studied beyond leading order except the case with $\pi \gamma^* \to \gamma$\cite{NLi} in which
the study is perform with Feynman gauge. Because the perturbative parts in the
factorization are extracted from scattering amplitudes of off-shell partons
and scattering amplitudes of off-shell partons are not gauge-invariant, it is not expected
that the perturbative parts are gauge-invariant.
In \cite{VKT} it has been pointed out that such a
factorization is gauge-dependent because the perturbative parts contain soft divergences at loop-level which
depend on gauges. It should be noted that in the case with $\pi \gamma^* \to \gamma$ the problem
of gauge invariance appears beyond tree-level, because gluons are exchanged at one- or higher loops.
Taking the case with $\pi \gamma^* \to \pi$, it is easy to show the problem at tree-level,
because at that level gluon-exchange happens.
\par
In the $k_T$-factorization, the form factor in Eq.(2) can be factorized in a similar way
as in Eq.(9):
\begin{equation}
F_\pi (Q) \sim \int dx d^2 \tilde k_{1\perp} d\tilde y d^2  k_{2\perp}
\phi (x, k_{1\perp}) \phi (y, k_{2\perp}) H(x,y,k_{1\perp},k_{2\perp}).
\end{equation}
The definition of the $k_T$-dependent wave functions can be found in \cite{VKT,TMD1,NLi}.
The perturbative part $H$ is extracted from scattering of off-shell partons.
Instead of parton momenta in Eq.(4) one has the momenta
for these off-shell partons:
\begin{eqnarray}
p_1^\mu =\left  (\bar x_0 P^+, 0, \vec p_\perp \right ), \ \ \
p_2^\mu = \left  ( x_0 P^+, 0, -\vec p_\perp \right ), \ \
k_1^\mu =\left  ( 0, \bar y_0 K^-, \vec k_\perp  \right ), \ \ \
\ k_2^\mu = \left  (0,  y_0 K^-, -\vec k_\perp \right ).
\end{eqnarray}
At leading order the same diagrams in Fig.1. give contributions to the form factor and hence
to $H$. In calculating these one makes the projection by replacing the spinor products like those
in Eq.(5) with:
\begin{equation}
u(p_1)\bar v(p_2) \to \gamma^- \gamma_5, \ \ \ \  v(k_1) \bar u(k_2) \to \gamma_5\gamma^+.
\end{equation}
We take $\mu=-$ as before. Then the contribution to the form factor from the $u$-quark is
only from Fig.1a.. If the $k_T$-factorization respects the gauge invariance, this contribution
must be gauge-invariant.
In Feynman gauge one finds the hard part from the $u$-quark:
\begin{equation}
 H(x,y,p_\perp,k_\perp) \vert_u  = \frac{1}{ \bar x \bar y Q^2 +(\vec p_\perp -\vec k_\perp )^2}.
\end{equation}
In deriving this result one has used the power counting: $\bar x Q \sim \bar y Q \sim p_\perp \sim k_\perp\sim \delta$
and only the leading term in $\delta$ has been taken into account. To extract $H$ one has also used the tree-level
result of $k_T$-dependent wave functions with the off-shell partons:
\begin{equation}
\phi(x, k_{1\perp}) \sim \delta(x-x_0) \delta^2 (\vec k_{1\perp} -\vec p_\perp), \ \ \ \ \
\phi(y, k_{2\perp}) \sim \delta(y-y_0) \delta^2 (\vec k_{2\perp} -\vec k_\perp).
\end{equation}
It should be noted that there is no problem of gauge-invariance for the tree-level
result of wave functions, because there is no gluon exchange.
\par
Now we calculate the hard part in an axial gauge fixed with an arbitrary vector $w$, i.e., $w\cdot G=0$.
The gluon propagator in this gauge reads:
\begin{equation}
\frac{-i}{q^2+i\varepsilon} \left [
   g^{\mu\nu} - \frac{ w^\mu q^\nu + w^\nu q^\mu}{w\cdot q} + w^2 \frac{q^\mu q^\nu}{(w\cdot q)^2} \right ].
\end{equation}
In the above the first term is just the propagator in Feynman gauge.
We obtain the contribution from the $u$-quark to the hard part in the axial gauge as:
\begin{eqnarray}
 H(x,y,p_\perp,k_\perp)\vert_u &=&  \frac{1}{ \bar x \bar y Q^2 +(\vec p_\perp -\vec k_\perp )^2}
   \left \{  1 -\frac{w_\perp \cdot (p-k)_\perp}{w\cdot (p_1-k_1)}
   + w^2 \frac{p_\perp \cdot (p-k)_\perp}{2 (w\cdot (p_1-k_1))^2}
\right.
\nonumber\\
   && \left. + \frac{1}{2 w \cdot (p_1-k_1) }\left [ \frac{l\cdot w}{\bar y K^-}
    k_\perp \cdot (k-p_\perp) + w\cdot k_\perp \right ] \right \}.
\end{eqnarray}
In this result the momentum fraction $x_0$ in $p_1$ and $y_0$ in $k_1$  in Eq. (61) should be
replaced with $x$ and $y$, respectively.
In the above $\{ \cdots \}$ every term is at order of ${\mathcal O}(\delta^0)$. Therefore
no term can be neglected with the power counting.
This result clearly indicates that $H$ is gauge-dependent. It is interesting to note that
the first two terms are canceled
and all terms in $H$ depend on $w$, if we take the vector $w$ as $w^\mu =(0,0,w_\perp)$.
For showing this clearly we give the explicit result for this case:
\begin{eqnarray}
 H(x,y,p_\perp,k_\perp)\vert_u &=&  \frac{1}{ \bar x \bar y Q^2 +(\vec p_\perp -\vec k_\perp )^2}
   \left \{
    \frac{w_\perp^2 p_\perp \cdot (p-k)_\perp}{2 (w_\perp\cdot (p_\perp-k_\perp))^2} +
  \frac{ w_\perp \cdot k_\perp}{2 w_\perp \cdot (p_\perp-k_\perp) }  \right \}.
\end{eqnarray}
\par
One can also show that the hard part is gauge-dependent in the general covariant gauge\cite{WZT}. In
the gauge the gluon propagator reads:
\begin{equation}
\frac{-i}{q^2+i\varepsilon} \left [
   g^{\mu\nu} - \alpha \frac{ q^\mu q^\nu}{q^2+i\varepsilon} \right ].
\end{equation}
In this gauge the contribution from the $u$-quark reads:
\begin{equation}
H(x,y,p_\perp,k_\perp)\vert_u = \frac{1}{ \bar x \bar y Q^2 +(\vec p_\perp -\vec k_\perp )^2}
  \left ( 1  + \frac{\alpha}{2}\frac{ p_\perp \cdot (p_\perp -k_\perp)}
  { \bar x \bar y Q^2 +(\vec p_\perp -\vec k_\perp )^2} \right ).
\end{equation}
Here the gauge-dependent term in $(\cdots)$ is proportional to $\alpha$. This term is at order
of ${\mathcal O}(\delta^0)$ in comparison with the first term. Therefore, the gauge-dependent term
can not be neglected. One notices here that the transverse momenta appear in the numerator. One may argue
that these terms with the transverse momenta in the numerator may be factorized with higher-twist
operators other than the leading-twist operator used to defined $\phi(x,p_\perp)$. Even if one can do so,
these terms contributing to the factorization hence to the form factor
are still gauge-dependent and can not be neglected
with the power counting in comparison with the term factorized with $\phi(x,p_\perp)$.
The conclusion here is that the $k_T$-factorization here at tree-level is gauge-dependent.
Beyond tree-level, the hard part will receive gauge-dependent contributions which are divergent\cite{VKT,FVKT}.

\par\vskip20pt
\par\vskip20pt
\par\noindent
{\bf\large Acknowledgments}
\par
This work is supported by National Nature Science Foundation of China(No. 10721063, 10575126 and 10975169).
F. Feng is supported by China Postdoctoral Science Foundation funded project. 
\par



\vskip20pt





\begin{thebibliography}{99}

\bibitem{BL} G.P. Lepage and S.J. Brodsky, Phys. Rev. D22 (1980) 2157.

\bibitem{CZrep} V.L. Chernyak and A. R. Zhitnitsky, Phys. Rept.  {\bf 112}
(1984) 173.

\bibitem{FGOC} R.D. Field, R. Gupta, S. Otto and L. Chang, Nucl. Phys. B186 (1981) 429.

\bibitem{DR} F.-M. Dittes and A.V. Radyushkin, Yad. Fiz. {\bf 34} (1981) 529, Sov. J. Nucl. Phys. {\bf 34}
(1981) 293.

\bibitem{KR} R.S. Khalmuradov and A.V. Radyushkin, Yad. Fiz. {\bf 42} (1985) 458, Sov. J. Phys. {\bf 42} (1985) 289.

\bibitem{BT} E. Braaten and S.-M. Tse, Phys. Rev. D35 (1987) 3255.

\bibitem{BNP} B. Melic, B. Nizic and K. Passek, Phys. Rev. D60 (1999) 074004,
e-Print: hep-ph/9802204.

\bibitem{CS} J.C. Collins and D.E. Soper, Nucl. Phys. B193 (1981) 381[Erratum-ibid.
B213 (1983) 545], Nucl. Phys. B197 (1982) 446.

\bibitem{CSS} J.C. Collins, D.E. Soper and G. Sterman, Nucl. Phys. B250 (1985) 199.

\bibitem{JMY} X.D. Ji, J.P. Ma and F. Yuan, Phys. Rev. D71 (2005) 034005, Phys. Lett. B597 (2004) 299.

\bibitem{JMYG} X.D. Ji, J.P. Ma and F. Yuan,  JHEP 0507:020,2005, hep-ph/0503015.

\bibitem{FMW} F. Feng, J.P. Ma and Q. Wang, JHEP 0706 (2007) 039,
e-Print: arXiv:0704.3782 [hep-ph]

\bibitem{Ster} G. Sterman, Nucl. Phys. B281 (1987) 310.

\bibitem{LS} H.-n. Li and G. Sterman, Nucl. Phys. B381 (1992) 129.

\bibitem{VKT} F. Feng, J.P. Ma and Q. Wang, Phys. Lett. B674 (2009) 176,
e-Print: arXiv:0807.0296 [hep-ph], e-Print: arXiv:0901.2965 [hep-ph].

\bibitem{FVKT} F. Feng, J.P. Ma and Q. Wang, Phys. Lett. B677 (2009)121-122,
e-Print: arXiv:0808.4017 [hep-ph], H.-nan Li and S. Mishima, Phys.Lett. B674 (2009) 182,
e-Print: arXiv:0808.1526 [hep-ph].


\bibitem{TMD1} J.P. Ma and Q. Wang, Phys.Rev. D75 (2007) 014014,
hep-ph/0607234, Phys.Lett. B642 (2006) 232, hep-ph/0605075,

\bibitem{TMDB} J.P. Ma and Q. Wang, JHEP 0601 (2006) 067, hep-ph/0510336, Phys.Lett. B613 (2005) 39,
hep-ph/0412282.

\bibitem{JH} J.C. Collins and F. Hautmann, Phys. Lett. B472 (2000)129,
e-Print: hep-ph/9908467

\bibitem{ASY} R. Akhoury, G. Sterman and Y.-P. Yao, Phys. Rev. D50 (1994) 358.

\bibitem{BrFi} V.M. Braun and I.E. Filyanov, Z. Phys. C44 (1989) 157, see also C48 (1990) 239.

\bibitem{Exp}  C.J. Bebek {\it et al.}, Phys.Rev. D 17 (1978)  1693.

\bibitem{NLi} S. Nandi and H.-n. Li, Phys. Rev. D76 (2007) 034008.

\bibitem{WZT} Z.T. Wei, private communication.



\end{thebibliography}
\end{document}